# Acousto-electric study of microwave-induced current domains

B. Friess[1], I. A. Dmitriev[2,3], V. Umansky[4], L. Pfeiffer[5], K. West[5], K. von Klitzing[1] and J. H. Smet[1]

**Affiliations**

[1] Max Planck Institute for Solid State Research, Heisenbergstrasse 1, D-70569 Stuttgart, Germany

[2] Department of Physics, University of Regensburg, 93040 Regensburg, Germany

[3] Ioffe Physical Technical Institute, 194021 St. Petersburg, Russia

[4] Braun Centre for Semiconductor Research, Department of Condensed Matter Physics, Weizmann Institute of Science, Rehovot 76100, Israel

[5] Department of Electrical Engineering, Princeton University, Princeton, New Jersey 08544, USA

**Abstract**

Surface Acoustic Waves (SAW) have been utilized to investigate the properties of a two-dimensional electron system, subjected to a perpendicular magnetic field and monochromatic microwave radiation, in the regime where the so-called microwave-induced zero-resistance states form. Contrary to conventional magneto-transport in Hall bar and van der Pauw geometries, the collimated SAW beam probes only the bulk of the electronic system exposed to this wave. Clear signatures appear in the SAW propagation velocity corroborating that neither contacts, nor sample edges are a root source for their emergence. By virtue of the directional nature of this probing method and with the assistance of theoretical modelling, we were also able to demonstrate that the SAW response depends on the angle between its propagation vector and the orientation of domains which spontaneously form when zero-resistance is observed in transport. This confirms in unprecedented manner the formation of an inhomogeneous phase under these non-equilibrium conditions.

Driving a two-dimensional electron system exposed to a perpendicular magnetic field out of equilibrium by applying monochromatic microwave radiation with frequency $f$ adds the photon energy $hf$ as an additional quantized energy scale to the system. In samples of sufficient quality, an additional set of magnetoresistance oscillations appears, usually prior to the full development of Shubnikov-de-Haas oscillations [1,2]. These microwave-induced oscillations display minima (maxima) at magnetic field values below (above) the nodal points where the photon energy equals an integer multiple of the cyclotron energy, $hf = i \cdot hf_c$. Here $i$ is an integer, $hf_c$ the cyclotron energy and $h$ the Planck constant. If these oscillations become sufficiently strong, the resistance may tend to turn negative [3,4] and the electronic system enters the unstable regime forcing a spontaneous rearrangement of the electrons. The latter is believed to lead to a pattern of domains supporting alternating current flow such that the overall sum of the current equals the externally imposed current through the sample. The local longitudinal conductivity inside such a domain equals zero leading to a vanishing longitudinal resistance measured along the sample perimeter [5]. Since their discovery, both phenomena - microwave-induced resistance oscillations (MIRO) and zero-resistance states (ZRS) – have sparked intense efforts to gauge the origin of the underlying effects [6]. Different theoretical concepts have been put forward, that trace MIRO back to, among others, microwave-assisted impurity scattering (displacement mechanism) [7-10], a non-equilibrium energy distribution (inelastic mechanism) [10-12], classical memory effects [13-15], ponderomotive forces near contacts [16], and modifications to the edge transport [17]. The nature of the ZRS instability and properties of the domain state were addressed theoretically in Refs. [5,9,10,18-23]. Experimentally, major progress has been made in addressing key properties of MIRO and the zero-resistance states [6,24-27], however some basic questions still remain unanswered up to date [28,29,30].

In this article, surface acoustic waves serve as a complementary probe to investigate these microwave-induced resistance oscillations and the formation of current-domains in the zero-resistance regime. The propagation of sound is advantageous compared to electrical transport measurements for multiple reasons. It is a contactless technique. Sound propagation is directional and the probing direction is unaffected by the internal structure of the electronic system. This is particularly beneficial when investigating inhomogeneous or anisotropic phases that may for instance cause a complex spatial distribution of the electronic current flow difficult to detect or explore just from measuring voltage drops along the edges [31]. In our experimental setup, the surface acoustic waves (SAW) are generated and detected by interdigital transducers made from copper on top of the GaAs/AlGaAs heterostructure hosting the 2D electron system as shown in the inset to Fig. 1. When applying a high-frequency voltage to the transducers, a surface acoustic wave is excited in the piezo-electric substrate and propagates as a collimated beam towards the transducer on the opposite side of the sample. While traversing the mesa, the SAW interacts with the two-dimensional electron system that forms inside of a GaAs quantum well. Depending on the electron conductivity, the sound wave experiences a change of its propagation velocity v= $\omega/q$ (with $\omega$ as the SAW angular frequency and $q = 2\pi/\lambda$ the SAW wavevector) due to a partial screening of the SAW piezo-electric field. This minute adjustment of the sound velocity can be measured sensitively as a shift of the SAW phase $\Delta\alpha$ at a second transducer with the help of a vector network analyzer [32]. The change of the sound velocity $\Delta v = v - v_0$ can be calculated simply by the geometric expression $\Delta v/v = \lambda/L \cdot \Delta\alpha/360°$, where $\lambda$ denotes the SAW wavelength and L the interaction length (size of the 2DES). Both the SAW phase velocity and the SAW attenuation coefficient $\kappa$ are generally expressed through the longitudinal dielectric function $\varepsilon(q, \omega)$ as [33]

$$(v - v_0)/v - i\kappa/q = K_{\text{eff}}^2/2\varepsilon(q,\omega) \ . \qquad \text{Equation (1)}$$

In turn, $\varepsilon(q,\omega) = 1 + i\sigma_{xx}(q,\omega)/\sigma_m$ is determined by the linear longitudinal conductivity $\sigma_{xx}(q,\omega)$ along the direction of SAW propagation. In GaAs-based heterostructures, the electro-elastic coupling equals $K_{\text{eff}}^2 = 0.64 \cdot 10^{-3}$, with the parameter $\sigma_m = 2\varepsilon_0\varepsilon_m v = 0.33$ μS, where $\varepsilon_m \cong (12.5 + 1)/2 = 6.75$ is the arithmetic mean of the relative GaAs/AlGaAs and vacuum permittivities. The perfect screening case $\varepsilon \to \infty$ is historically taken as a reference and yields $v = v_0$ and $\kappa = 0$. To illuminate the sample with quasi-monochromatic microwave radiation, it is placed in the beam of a horn antenna at the end of a stainless steel hollow-core waveguide integrated into the sample rod of a top-loading-into-mixture dilution refrigerator system with superconducting coil. Studies of two different samples with similar structural design are presented in this publication. Both samples exhibit as single GaAs quantum well of width 29 nm (sample A) and 30 nm (sample B) in an AlGaAs host structure with double-sided doping, giving rise to electron densities of $3.0 \cdot 10^{11}$ cm$^{-2}$ (sample A) and $2.8 \cdot 10^{11}$ cm$^{-2}$ (sample B). Low-temperature mobilities are approximately 20-30×10$^6$ cm$^2$/Vs.

This SAW technique has been successfully deployed in the past in the context of 2D electron systems exposed to a magnetic field to explore changes in the density of states and compressibility as manifested in Shubnikov-de Haas oscillations and the quantum Hall effects [34,35], geometric resonances of electrons and composite fermion quasi-particles [36-38], as well as charge clustering and charge density wave physics [31,32]. As an example, Fig. 1 shows a typical trace of the longitudinal resistance and the sound velocity shift as a function of the magnetic field in the absence of microwave radiation. The phase velocity at zero magnetic field is taken as a reference: Δv=0. In this regime, electrons are capable to screen the piezo-electric field due to their large conductivity and the non-zero density of states of the electronic system at the chemical potential [34-36]. When the density of states at the chemical potential drops due to the formation of Landau levels and their successive depletion, the screening capability of the electrons reduces in a 1/B-periodic fashion. This is accompanied by a 1/B-periodic piezoelectric stiffening and an increase of the SAW propagation velocity. Therefore, the SAW signal is out of phase and displays a maximum whenever the magnetoresistance exhibits a minimum. We note that the resistive transport measurement, monitoring the potential drop along the sample boundary, differs from the SAW signal in the sense that the latter probes the conductivity along the propagation direction in the bulk area covered by the collimated SAW beam.

Fig. 2 illustrates the behavior in the presence of microwave radiation with a frequency *f* of 100 GHz. The top panel shows the magnetotransport data with well-pronounced MIRO as well as multiple zero-resistance states (for *i* = 1, 2 and 3). The sound propagation velocity is plotted in panel b and is strongly modulated as well, in particular at those magnetic fields where the microwave radiation induces zero-resistance states. The SAW power detected at the receiving transducer is shown in panel c. Here too, transmission features do not only appear at the fundamental ($i = 0$) but also at the location of the higher order zero-resistance states ($i = 2$ and $3$). The transmitted power increases in the field regions of zero resistance. Heating due to absorption at the cyclotron resonance can therefore be safely discarded as the root cause for the observed acoustic signal. Instead, the microwave-induced modulation of the electron conductivity in the bulk of the sample appears as a reasonable explanation. It causes a change in the piezoelectric restoring force, which in turn affects the sound velocity. Since the SAW propagation is confined to the bulk region covered by the collimated SAW beam, contact and edge transport phenomena can be excluded as an alternative explanation for microwave-induced zero-resistance states [16,17].

A closer inspection of the SAW velocity shift in Fig 2b reveals a slanted peak-like structure in the region of each zero-resistance state. However, during some cool-downs a fundamentally different SAW response was observed as illustrated in Fig. 3a. These data have been recorded at a microwave frequency of 50 GHz. Here, the SAW response consists of a minimum located near the center of the field interval covered by the zero-resistance state and separated from the remaining SAW response by two well-pronounced peaks that occur at the outskirts of the zero-resistance state where the resistance has started to rise rapidly. In the remainder, we address this important difference by modelling the expected response depending on the orientation of the current flow in the domains with respect to the propagation direction of the SAW.

As outlined in more detail in the Supplemental Material, in the range of magnetic fields relevant for MIRO the local dielectric function can be expressed as

$$\varepsilon(q = 0, \omega) = 1 + \omega_p^2/\omega_c^2 + i\sigma_{dc}(0)/\sigma_m \,. \qquad (2)$$

Here, $\sigma_{dc}(0) = \sigma_0(1 + \delta\sigma)$ is the linear dc conductivity as measured in MIRO transport experiments with $\delta\sigma$ being the oscillating microwave-induced conductivity contribution [6]. The term $\omega_p^2/\omega_c^2$ originates from the dissipationless imaginary part of $\sigma_{xx}(q, \omega)$. It is unaffected by the microwave illumination and it contains the 2D plasmon frequency at the SAW wave vector $q$, given by $\omega_p^2 = e^2 n_e q / 2m\varepsilon_0\varepsilon_m$. This description reproduces well the relation between the observed SAW velocity shift and MIRO in the regions outside of the zero-resistance states. In the latter regime, however, it fails because the local dc conductivity $\sigma_{dc}(0)$ becomes negative causing an instability and the formation of domains. It can be shown on general grounds that the local dissipative conductivity $\sigma_{dc}(E)$ should restore positive values in the limit of large static electric fields, $E=|\boldsymbol{E}|>E_d$, [5, 9, 12] and that the system state with $|\boldsymbol{E}|=E_d$ is stabilized [5]. This theory predicts the emergence of domains with a strong spontaneous electric field $\boldsymbol{E}_d$ and a non-dissipative Hall current $\boldsymbol{j}_H = en(\boldsymbol{B} \times \boldsymbol{E}_d)/B^2$ flowing in the direction perpendicular to $\boldsymbol{E}_d$, such that the local dissipative current, $\boldsymbol{j}_d = \boldsymbol{E}\sigma_{dc}(E)$, as well as the net resistance measured in transport experiments vanish.

As long as the amplitude of the SAW field is smaller than $E_d$, the local screening properties in the domains of the zero-resistance state are determined by the conductivity linearized around $E_d$. The SAW response is then described by Eq. (2), but with $\sigma_{dc}(0)$ replaced by the differential conductivity $\sigma_{diff} = \partial(j_d)_x/\partial E_x|_{E=E_d}$ with the derivative taken along the direction of the SAW electric field, $x$. This differential conductivity reduces to

$$\sigma_{\text{diff}} = E_d \cos^2 \theta_d\, \sigma'_{dc}(E_d), \qquad \text{Equation (3)}$$

where $\theta_d$ is the local angle between the domain and the SAW field and $\sigma'_{dc}(E) = d\sigma_{dc}(E)/dE$. In full accord with our observations, the above description predicts a kink in the dependence $\Delta v(B)$ at each transition to a zero-resistance state where the linear conductivity $\sigma_{dc}(0)$ turns negative while $\sigma_{\text{diff}} \geq 0$, describing screening in the domain state, remains non-negative. In the case of $\cos\theta_d = 0$, when the SAW field is perpendicular to the domain field, i.e. collinear with the Hall current in the domain, $\sigma_{\text{diff}}$ vanishes, and the dielectric function reduces to $1 + \omega_p^2/\omega_c^2$, resulting in an apparent linear behavior of $\Delta v(B)$ in the region of ZRS as seen in Fig. 3b. For $\cos^2\theta_d > 0$, the above theory predicts minima in $\Delta v(B)$ inside the ZRS regions, as opposed to maxima of $\Delta v(B)$ at those MIRO minima where $\sigma_{dc}(0)$ remains positive. This behavior is highlighted in Fig. 3c. For illustrative purposes, we use $\sigma_{dc}(E) = \sigma_0[1 + \delta\sigma/(1 + E^2/E_0^2)]$, in which case $\sigma_{\text{diff}} = 2\cos^2\theta_d\,(1 - 1/\delta\sigma)$. This result indeed closely resembles the experimental trace in Fig. 3a. The take home message is that the directional response of the surface acoustic waves can be used to probe the orientation of the current domains. The dissimilarity of the measurements in Fig. 2b and 3a reflect a different alignment of the current domains.

In summary, surface acoustic waves were deployed to probe state-of-the-art two-dimensional electron systems irradiated by microwaves. The directional nature of these surface waves enables to probe the conductivity in the bulk of the sample in the MIRO and zero-resistance regimes rather than along the sample boundary, as is the case in conventional transport studies. The clear signatures for microwave-induced zero-resistance states in the propagation velocity of the SAW offers unequivocal evidence that this phenomenon is neither an edge nor a contact effect. Theoretical modelling, assuming the formation of current domains in the zero-resistance regime and including the angle between the SAW propagation direction and the orientation of the domains as a parameter, qualitatively reproduces the two distinct SAW responses observed in experiment, thereby substantiating that directional probing with SAW is a complementary and powerful tool to probe inhomogeneous phases that may form in 2D electron systems under non-equilibrium conditions.

Acknowledgments:

This work was supported by the German-Israeli Foundation for Scientific Research and Development (GIF). I.A.D acknowledges financial support from the Deutsche Forschungsgemeinschaft (project DM1/4-1). The work at Princeton University was funded by the Gordon and Betty Moore Foundation through the EPiQS initiative Grant GBMF4420, and by the National Science Foundation MRSEC Grant DMR 1420541.

Figures:

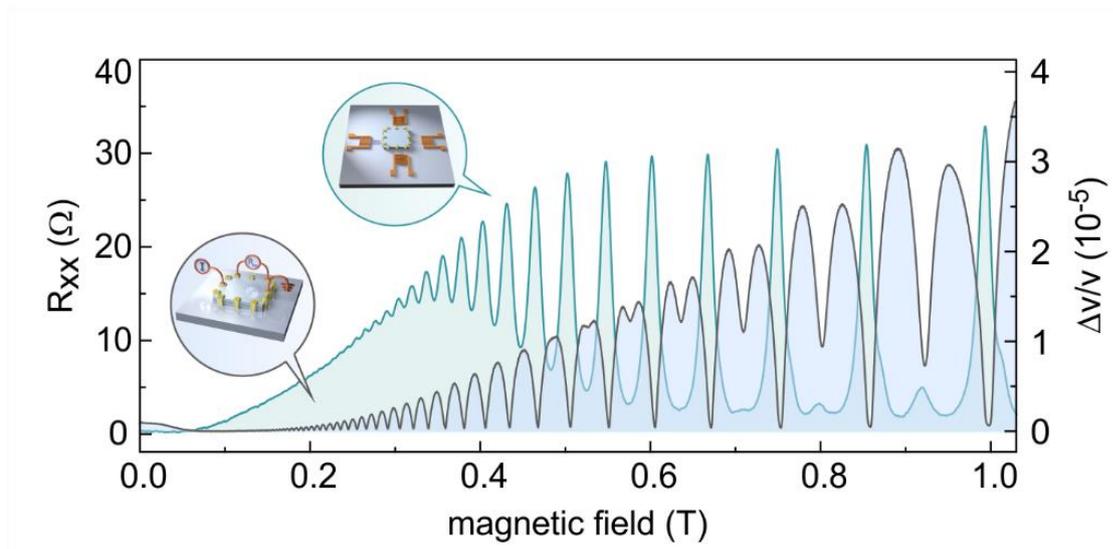

Figure 1: Longitudinal resistance and SAW velocity shift as a function of applied magnetic field recorded on sample A with an electron density of $3.0 \cdot 10^{11}$ cm$^{-2}$.

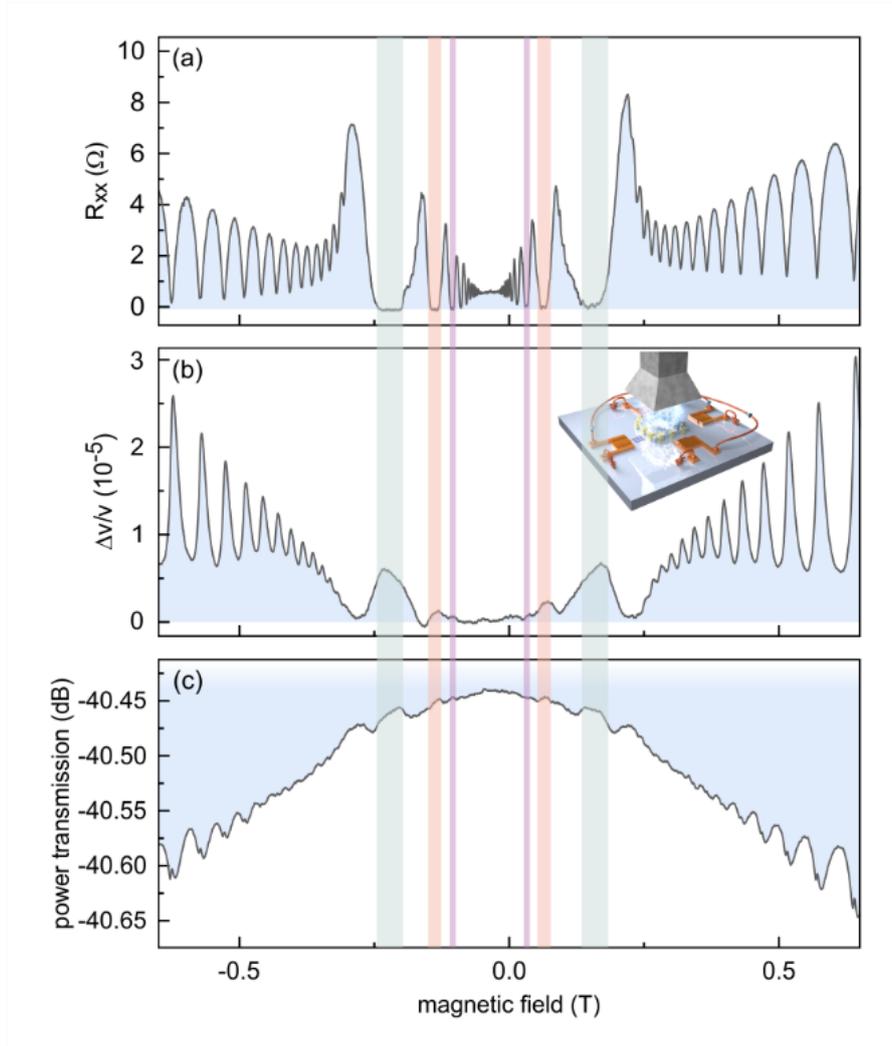

Figure 2: Microwave-induced resistance oscillations and zero-resistance states probed by surface acoustic waves on sample A. (a) Longitudinal resistance as a function of magnetic field under microwave exposure (frequency $f$=100 GHz). Colored bars mark the three zero-resistance states. The general shift of the data with respect to the x-axis is due to the magnetic field offset of the magnet. (b) Velocity shift of surface acoustic waves measured concomitantly with the resistive transport measurements in panel a. The inset shows the sample setup with interdigital transducers placed aside the square-shaped mesa with the two-dimensional electron system underneath the horn of the microwave hollow-core waveguide. (c) Transmitted SAW power detected with the receiving transducer after traversing the mesa.

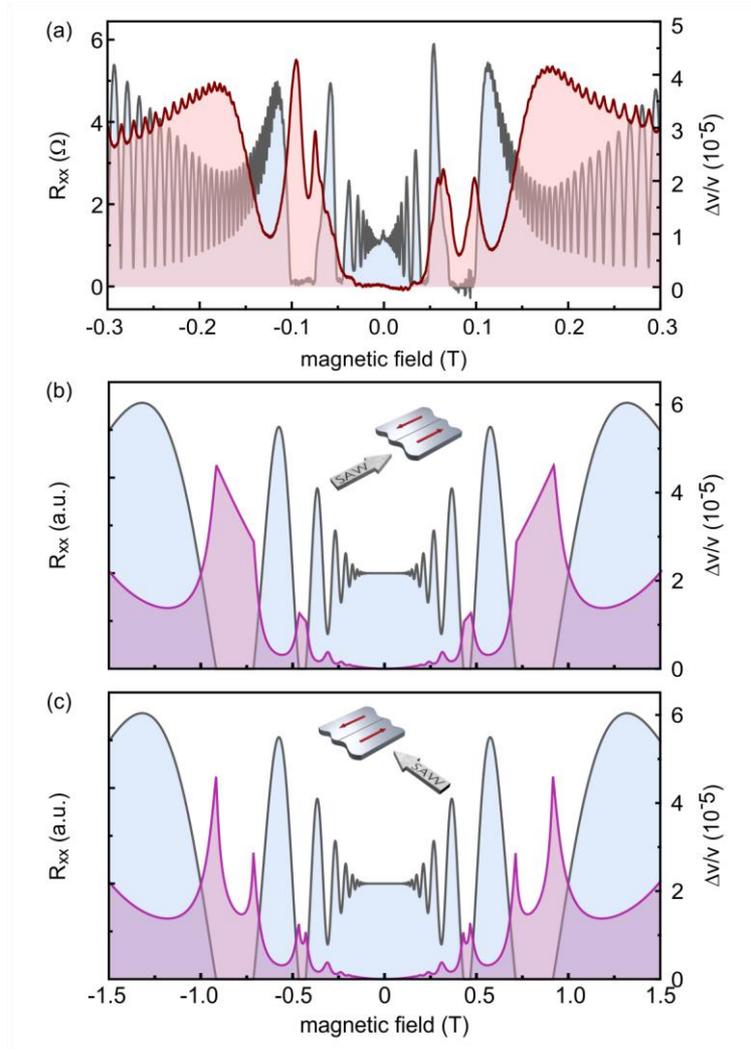

Figure 3: (a) Experimental example of the second type of surface acoustic wave response in the zero-resistance regime. This measurement was recorded for a microwave frequency $f$ =50 GHz on sample B. Plotted is the longitudinal resistance (blue) and SAW velocity shift (red). (b) and (c) Calculated magnetoresistance (blue) and SAW velocity shift (pink) for the SAW propagation direction along [panel (b)] and perpendicular [panel (c)] to the current in the domains.